\documentclass[12pt]{article}
\usepackage{graphicx}

\def\beq{\begin{equation}}
\def\eeq{\end{equation}}
\def\bea{\begin{eqnarray}}
\def\eea{\end{eqnarray}}
\def\nn{\nonumber}
\def\sss{\scriptscriptstyle}

\def\barp{{\raise.35ex\hbox
{${\sss (}$}}---{\raise.35ex\hbox{${\sss )}$}}}
\def\bdbarp{\hbox{$B_d$\kern-1.4em\raise1.4ex\hbox{\barp}}}
\def\bsbarp{\hbox{$B_s$\kern-1.4em\raise1.4ex\hbox{\barp}}}

\def\barpk{{\raise.35ex\hbox
{${\sss (}$}}--{\raise.35ex\hbox{${\sss )}$}}}
\def\kbarp{\hbox{$K$\kern-0.9em\raise1.4ex\hbox{\barpk}}}
\def\roughly#1{\mathrel{\raise.3ex\hbox
{$#1$\kern-.75em\lower1ex\hbox{$\sim$}}}}

\def\gsim{\roughly>}

\def\adir00{{a_{\sss dir}^{00}}}

\def\B00{B^{00}}
\def\Bp0{B^{+0}}

\def\dsp{\displaystyle}

\def\npb#1#2#3{{ Nucl.\ Phys.} {\bf B#1}, #3 (#2)}
\def\plb#1#2#3{{ Phys.\ Lett.} {\bf #1B}, #3 (#2)}
\def\prd#1#2#3{{ Phys.\ Rev.} {\bf D#1}, #3 (#2)}
\def\newprd#1#2#3{{ Phys.\ Rev.} {\bf D#1}, #3 (#2)}

\def\zpc#1#2#3{{ Zeit.\ Phys.} {\bf C#1}, #3 (#2)}

\begin{document}

\setlength{\baselineskip}{20pt}

\begin{flushright}
UdeM-GPP-TH-02-107 \\
IMSc-2002/09/33 \\
\end{flushright}

\begin{center}
\bigskip
{\Large \bf Lepton Polarization and Forward-Backward Asymmetries in $b
  \to s \tau^+ \tau^-$} \\
\bigskip
Wafia Bensalam $^{a,}$\footnote{wafia@lps.umontreal.ca},~~ David
London $^{a,}$\footnote{london@lps.umontreal.ca},~~ Nita Sinha
$^{b,}$\footnote{nita@imsc.res.in},~~ Rahul Sinha
$^{b,}$\footnote{sinha@imsc.res.in} \\

\end{center}


\begin{flushleft}
~~~~~~~~~~~$a$: {\it Laboratoire Ren\'e J.-A. L\'evesque, 
Universit\'e de Montr\'eal,}\\
~~~~~~~~~~~~~~~{\it C.P. 6128, succ. centre-ville, Montr\'eal, QC,
Canada H3C 3J7}\\
~~~~~~~~~~~$b$: {\it Institute of Mathematical Sciences, Taramani,
 Chennai 600113, India}
\end{flushleft}

\begin{center} 
\bigskip (\today)
\vskip0.5cm
{\Large Abstract\\}
\vskip3truemm
\parbox[t]{\textwidth} {We study the spin polarizations of both $\tau$
  leptons in the decay $b\to s\tau^+\tau^-$. In addition to the
  polarization asymmetries involving a single $\tau$, we construct
  asymmetries for the case where both polarizations are simultaneously
  measured. We also study forward-backward asymmetries with polarized
  $\tau$'s. We find that a large number of asymmetries are predicted
  to be large, $\gsim 10\%$. This permits the measurement of all
  Wilson coefficients and the $b$-quark mass, thus allowing the
  standard model (SM) to be exhaustively tested. Furthermore, there
  are many unique signals for the presence of new physics. For
  example, asymmetries involving triple-product correlations are
  predicted to be tiny within the SM, $O(10^{-2})$. Their observation
  would be a clear signal of new physics.}
\end{center}
\thispagestyle{empty}
\newpage
\setcounter{page}{1}
\textheight 21.0 true cm
\baselineskip=14pt

\section{Introduction}

There has been a great deal of theoretical work examining the decay $b
\to s \ell^+ \ell^-$, both at the inclusive and exclusive level
\cite{bsllSM}. As usual, the hope is that, through precision
measurements of this decay, one will find evidence for the presence of
physics beyond the standard model (SM). Indeed, this decay mode has
been extensively studied in various models of new physics
\cite{bsllNP}.

Some years ago, it was noted that the measurement of the polarization
of the final-state $\tau^-$ in the inclusive decay $b \to X_s \tau^+
\tau^-$ can provide important information about the Wilson
coefficients of the underlying effective Hamiltonian
\cite{AGM,Hewett,KS}. Within the SM, this inclusive decay is described
in terms of five theoretical parameters: the four Wilson coefficients
($C_7$, $C_{10}$ and real and imaginary parts of $C_9$), and the mass
of the $b$-quark, $m_b$. In principle, all of these theoretical
parameters can be completely determined using measurements of the
three $\tau^-$ polarization asymmetries, the total (unpolarized) rate,
and the forward-backward (FB) asymmetry.

In practice, however, the SM $\tau^-$ polarization asymmetry along the
normal component is expected to be $O(10^{-2})$ \cite{smallPNref}, and
is therefore probably too small to be measured. This situation can be
remedied to some extent if, in addition to the polarization
asymmetries of the $\tau^-$, we also consider similar asymmetries for
the $\tau^+$ \cite{FKY}. This adds one more independent observable.
However, even if the sizeable polarization asymmetries of both
$\tau^+$ and $\tau^-$ can be separately measured, there are only as
many measurements as there are unknowns, so that there are no
redundant measurements to provide crosschecks for the SM. Furthermore,
this program requires that the flavor of the $b$-quark be tagged: in
an untagged sample, there are only four observables, since the
measurement of the FB asymmetry requires tagging. It will therefore be
very difficult to rigorously test the SM if only single
$\tau$-polarization measurements are made in $b \to X_s \tau^+
\tau^-$.

In this paper, we try to construct the maximum possible number of
independent observables. This is achieved by considering the situation
in which both $\tau^+$ and $\tau^-$ polarizations are simultaneously
measured. As we will see, a variety of new asymmetries can be
constructed in this case. We compute the polarization and
forward-backward asymmetries for both singly-polarized and
doubly-polarized final-state leptons. A large number of these new
asymmetries do not require the tagging of the $b$-quark. (Note that,
in an untagged sample, while the FB asymmetry for unpolarized leptons
vanishes, some of the FB asymmetries for polarized leptons are
nonvanishing.) On the other hand, if $b$-tagging is possible, the
measurement of these new asymmetries provides even more information.
The polarized FB asymmetries as well as the double-spin polarization
asymmetries all depend in different ways on the Wilson coefficients,
so that these coefficients can be obtained in many different
ways. This redundancy provides a huge number of crosschecks, and
allows the SM to be exhaustively tested. An interesting consequence of
the large number of observables, is that $m_b$ can be extracted. If
the phenomenologically-obtained value of $m_b$ were to agree with
theoretical estimates \cite{luke}, this would be an important step in
confirming our understanding of QCD,

In our calculations we consider only contributions from SM operators.
However, using arguments based on CPT invariance and the properties of
the SM operators under C, P and T, we derive relations between these
observables which are clean tests of new physics. Some of these tests
rely on the fact that within the SM there are negligible CP-violating
contributions to the decay mode being considered. Our philosophy is to
test for the presence of new physics (NP) without considering the
detailed structure of the various operators that can contribute to
NP. Should a signal for NP be seen, the consideration of specific NP
operators would help in determining the nature of NP contributions
(for example, see Ref.~\cite{FKY}).

We begin in Sec.~2 with a discussion of the calculation of $|{\cal
M}|^2$, where ${\cal M}$ is the amplitude for $b \to s \tau^+ \tau^-$
(the results of the calculation of ${\cal M}$ are complicated, and are
presented in the Appendix). The polarization asymmetries and
forward-backward asymmetries are examined in Secs.~3 and 4,
respectively. We discuss these asymmetry measurements within a variety
of scenarios in Sec.~5. We conclude in Sec.~6.

In total, we numerically evaluate the differential decay rate and 31
asymmetries as a function of the invariant lepton mass. Note that it
will be extremely difficult to measure asymmetries smaller than 10\%,
as they would require $\sim 10^{10}$ $B$ mesons for a $3\sigma$ signal
(not including efficiencies for spin-polarization measurements and for
tagging). We therefore consider only asymmetries larger than 10\% as
measurable. If one can only measure an individual $\tau^+$ or $\tau^-$
spin, but cannot tag the flavor of the $b$-quark, then there are only
two sizeable observables. If $b$-tagging can be done and one can
measure the spin of the $\tau^+$ or $\tau^-$, this increases to 6
measurable asymmetries. However, if the polarizations of both the
$\tau$-leptons can be measured and flavor tagging of the $b$ is
possible, we find that nine of the asymmetries constructed here are
large in the SM. Including the decay rate, this leads to 10 sizeable
observables, which allows for a redundant test of the SM.

In addition, we find that the violation of certain SM asymmetry
relations are clean tests of NP. Some of these relations are violated
only in the presence of CP-violating NP. A large numbers of these
asymmetries are $O(10^{-2})$ in the SM, so that the observation of
larger asymmetries would be signals of NP. Certain combinations of
these asymmetries are identically zero in the SM, and hence are litmus
tests of NP.

\newpage

\section{$|{\cal M}|^2$ for $b \to s \tau^+ \tau^-$}

We begin by considering the calculation of $|{\cal M}|^2$ for $b \to s
\tau^+ \tau^-$. Including QCD corrections, the effective Hamiltonian
describing the decay $b \to s \tau^+ \tau^-$ \cite{hamiltonian} leads
to the matrix element
\beq
{\cal M} = T_9 + T_{10} + T_7 ~,
\label{amplitude}
\eeq
where
\beq
T_9  =  {\alpha G_F \over \sqrt{2} \pi} C_9^{eff}  V_{tb} V^*_{ts} \,
\Big[ {\bar s}_L \gamma_\mu b_L \Big] 
\Big[ {\bar \tau^{-}}  \gamma^\mu \tau^{+} \Big] ,
\eeq
\beq
T_{10}  =  {\alpha G_F \over \sqrt{2} \pi} C_{10}  V_{tb} V^*_{ts} 
 \Big[ {\bar s}_L \gamma_\mu b_L \Big] 
 \Big[ {\bar \tau^{-}}  \gamma^\mu \gamma_5 \tau^{+} \Big] ,
\eeq
\beq
T_7  =  {\alpha G_F \over \sqrt{2} \pi} C_7^{eff}  V_{tb} V^*_{ts}
 \left[ { -2i m_b \over q^2}  \right] 
 \Big[{\bar s}_L \sigma_{\mu\nu} q^{\nu} b_R \Big]
 \Big[ {\bar \tau^{-}}  \gamma^\mu \tau^{+} \Big] ~.
\eeq
In the above, $q$ is the momentum transferred to the lepton pair, and
we have neglected the $s$-quark mass. The Wilson coefficients $C_i$
are evaluated perturbatively at the electroweak scale and then evolved
down to the renormalization scale $\mu$. The coefficients $C_7^{eff}$
and $C_{10}$ are real, and take the values
\beq
C_7^{eff} = -0.315~,~~~~ C_{10} = -4.642 
\eeq
in the leading-logarithm approximation \cite{KS}. On the other hand,
the coefficient $C_9^{eff}$ is complex, and its value is a function of
$\hat{s} \equiv q^2/m_b^2$: $C_9^{eff}(\mu) \equiv
C_9(\mu)+Y(\mu,\hat{s})$, where the function $Y(\mu,\hat{s})$ contains
the one-loop contributions of the four-quark operators \cite{AGM,
hamiltonian}. An additional contribution to $C_9^{eff}$ arises from
the long-distance effects associated with real $c\bar{c}$ resonances
in the intermediate states \cite{resonances}. Thus, within the SM, the
decay $b \to s \tau^+ \tau^-$ is described by four Wilson coefficients
for a given value of $\hat{s}$: $C_7^{eff}$, $C_{10}$, ${\rm
Re}(C_9^{eff})$ and ${\rm Im}(C_9^{eff})$.

Because the expressions in $|{\cal M}|^2$ are complicated, we present
the actual results of this calculation in the Appendix. Note that the
polarization and forward-backward asymmetries, which will be discussed
in subsequent sections, are calculated as functions of the terms of
$|{\cal M}|^2$. Also, some signals of new physics are derived using
the C, P and T properties of the terms at the $|{\cal M}|^2$ level.

There is one point which is worth mentioning here. In the calculation
of $|{\cal M}|^2$, there are terms which involve the imaginary pieces
of the Wilson coefficients [e.g.\ the ${\rm Im}(C_9^{eff}C_{10}^*)$
term in Eq.~(\ref{T9T10})]. These are the coefficients of terms like
${\epsilon}_{\mu\alpha\beta\phi} \, p_s^\mu p_-^\alpha s_-^\beta
p_+^\phi$ in $|{\cal M}|^2$, which give rise to triple-product
correlations (e.g.\ ${\vec p}_- \cdot ({\vec p}_+ \times {\vec
s}_-)$). Naively, these triple products appear to violate
time-reversal symmetry (T) and so, by the CPT theorem, should also be
signals of CP violation. However, all the amplitudes in
Eq.~(\ref{amplitude}) have the same weak phase (neglecting the small
$u$-quark contribution in the loop), so that their interference cannot
give rise to CP violation. Thus, there appears to be an inconsistency.

What is happening is the following: a triple product is not a true
T-violating signal, since the action of T exchanges the initial and
final states. Because of this, triple-product correlations can be
faked by the presence of strong phases, even if there is no CP
violation. This is the situation which arises here -- nonzero strong
phases of the Wilson coefficients can lead to triple products.
Usually, it is CP violation which interests us, and we wish to
eliminate such fake signals. However, in this case, we are interested
in measuring the imaginary parts of the Wilson coefficients in order
to test the SM, so that these fake signals will be quite useful.

\section{Polarization Asymmetries}

In the computation of the various polarization asymmetries we choose a
frame of reference in which the leptons move back to back along the
$z$-axis, with the $\tau^-$ moving in the direction $+\widehat{z}$.
The $s$-quark then goes in the same direction as the $b$-quark, with
the $s$-quark making an angle $\theta$ with the $\tau^-$. Our specific
choices for the 4-momenta components are as follows:
\bea
  p^\mu_{\tau^-} &=& \{\sqrt{P^2+m_\tau^2},0,0,P\} ~,\nn\\
  p^\mu_{\tau^+} &=& \{\sqrt{P^2+m_\tau^2},0,0,-P\} ~,\nn\\
  p^\mu_s &=&\{K,0,K \sin\theta,K \cos\theta \} ~,\nn\\
  p^\mu_b &=&\{\sqrt{K^2+m_b^2},0,K\sin\theta,K \cos\theta \}.
\label{frame}
\eea

Using the above calculation of ${|\cal M|}^2$, we can compute the
decay rate for unpolarized leptons by summing over the lepton spins
and integrating over the angular variables. As a function of the
invariant mass of the lepton pair, this decay rate is given by
\beq
\left( \frac{d\Gamma(\hat{s})}{d\hat{s}}\right)_{\rm unpol} =
\frac{G_F^2\,m_b^5}{192\,\pi^3} \frac{\alpha^2}{4\,\pi^2} \left|
V_{tb}\,V_{ts}^* \right|^2 \, (1-\hat{s})^2 \, \sqrt{1 -
\frac{4\,{{\hat{m_\tau}}}^2}{\hat{s}}}\,\Delta ~,
\eeq
where $\hat{m_\tau} \equiv m_\tau/m_b$, and
\bea
\Delta&=&\Big( 12\,{\rm Re}(C_7^{eff}C_9^{eff^*})+
              \frac{4\,|C_7^{eff}|^2\,( 2 + \hat{s})}{\hat{s}}\Big)\,
              \Big( 1 + \frac{2\,\hat{m}_\tau^2}{\hat{s}} \Big) \\
   && + (|C_9^{eff}|^2+|C_{10}|^2)\,\Big( 1 + 2\,{\hat{s}} + 
       \frac{2\,( 1 - {\hat{s}}) \,\hat{m}_\tau^2}{\hat{s}}\Big)+6\,
       (|C_9^{eff}|^2-|C_{10}|^2)\,\hat{m}_\tau^2. \nn
\eea
This agrees with the earlier results
\cite{AGM,Hewett,KS,hamiltonian,Ali} in the appropriate limits.

We now consider the possibility that the polarizations of the
final-state leptons can be measured. The spins of the $\tau^\pm$ are
defined in their rest frames to be:
\beq
\widehat{s}^\mu_{\tau^-} = \left\{ 0,s^{-}_x,s^{-}_y,s^{-}_z \right\}
~~,~~~~
\widehat{s}^\mu_{\tau^+} = \left\{ 0,s^{+}_x,s^{+}_y,s^{+}_z \right\}
~.
\eeq
One can obtain the spins of the $\tau^\pm$ in the frame of
Eq.~(\ref{frame}) straightforwardly by performing a Lorentz boost:
\beq
s^\mu_{\tau^-} = \left\{ {P\over m_\tau} s^-_z , s^{-}_x , s^{-}_y,
{\sqrt{P^2 + m_\tau^2} \over m_\tau} s^{-}_z \right\} ~~,~~~~
s^\mu_{\tau^+} = \left\{ -{P\over m_\tau} s^+_z , s^+_x , s^{+}_y,
{\sqrt{P^2 + m_\tau^2} \over m_\tau} s^{+}_z \right\} ~.
\eeq

We now define differential decay rate as a function of the spin
directions of the $\tau^\pm$, ${\bf s^{+}}$ and ${\bf s^{-}}$, where
${\bf s^{+}}$ and ${\bf s^{-}}$ are unit vectors in the $\tau^\pm$
rest frames. This is given by
\bea
\frac{d\Gamma({\bf s^{+}},{\bf s^{-}},\hat{s})}{d\hat{s}} &=&
\frac{1}{4} \Bigg( \frac{d\Gamma(\hat{s})}{d\hat{s}}\Bigg)_{\rm
  unpol}\,\Bigg[1+\Bigg({\cal P}^{-}_x \, s_x^-
  +{\cal P}^{-}_y \, s_y^- +{\cal P}^{-}_z \, s_z^- \nn\\
&&\hskip2.2truein +~{\cal P}^{+}_x \,
  s_x^+ +{\cal P}^{+}_y \, s_y^+
  +{\cal P}^{+}_z \, s_z^+\Bigg) \nn \\
 && +\Bigg({\cal P}_{xx} \, s_x^+ s_x^- +{\cal P}_{xy}
  \, s_x^+ s_y^- +{\cal P}_{xz} \, s_x^+ s_z^- +{\cal P}_{yx} \, s_y^+
  s_x^- +{\cal P}_{yy} \, s_y^+ s_y^- \nn\\
&& \hskip0.5truein +~{\cal P}_{yz} \, s_y^+ s_z^-
  +{\cal P}_{zx} \, s_z^+ s_x^- +{\cal P}_{zy} \, s_z^+ s_y^- +{\cal
  P}_{zz} \, s_z^+ s_z^-
\Bigg)\Bigg],
\label{spinrate}
\eea
where the single-lepton polarization asymmetries ${\cal
P}_i^\mp(i=x,y,z)$ are obtained by evaluating
\bea
{\cal P}_i^- &=& \dsp\frac{\Big[\frac{d\Gamma({\bf s^{-}}={\bf
        \hat{i}},~{\bf s^{+}}={\bf
        \hat{i}})}{d\hat{s}}+\frac{d\Gamma({\bf s^{-}}={\bf
        \hat{i}},~{\bf s^{+}}={-\bf \hat{i}})}{d\hat{s}}\Big]
  -\Big[\frac{d\Gamma({\bf s^{-}}={-\bf \hat{i}},~{\bf s^{+}}={\bf
        \hat{i}})}{d\hat{s}}+\frac{d\Gamma({\bf s^{-}}={-\bf
        \hat{i}},~{\bf s^{+}}={-\bf \hat{i}})}{d\hat{s}}\Big]}
{\Big[\frac{d\Gamma({\bf s^{-}}={\bf \hat{i}},~{\bf s^{+}}={\bf
        \hat{i}})}{d\hat{s}}+\frac{d\Gamma({\bf s^{-}}={\bf
        \hat{i}},~{\bf s^{+}}={-\bf \hat{i}})}{d\hat{s}}\Big]
  +\Big[\frac{d\Gamma({\bf s^{-}}={-\bf \hat{i}},~{\bf s^{+}}={\bf
        \hat{i}})}{d\hat{s}}+\frac{d\Gamma({\bf s^{-}}={-\bf
        \hat{i}},~{\bf s^{+}}={-\bf \hat{i}})}{d\hat{s}}\Big]} ~,\nn\\
{\cal P}_i^+ &=&\frac{\Big[\frac{d\Gamma({\bf s^{-}}={\bf
        \hat{i}},~{\bf s^{+}}={\bf
        \hat{i}})}{d\hat{s}}+\frac{d\Gamma({\bf s^{-}}={-\bf
        \hat{i}},~{\bf s^{+}}={\bf \hat{i}})}{d\hat{s}}\Big]
  -\Big[\frac{d\Gamma({\bf s^{-}}={\bf \hat{i}},~{\bf s^{+}}={-\bf
        \hat{i}})}{d\hat{s}}+\frac{d\Gamma({\bf s^{-}}={-\bf
        \hat{i}},~{\bf s^{+}}={-\bf \hat{i}})}{d\hat{s}}\Big]}
{\Big[\frac{d\Gamma({\bf s^{-}}={\bf \hat{i}},~{\bf s^{+}}={\bf
        \hat{i}})}{d\hat{s}}+\frac{d\Gamma({\bf s^{-}}={-\bf
        \hat{i}},~{\bf s^{+}}={\bf \hat{i}})}{d\hat{s}}\Big]
  +\Big[\frac{d\Gamma({\bf s^{-}}={\bf \hat{i}},~{\bf s^{+}}={-\bf
        \hat{i}})}{d\hat{s}}+\frac{d\Gamma({\bf s^{-}}={-\bf
        \hat{i}},~{\bf s^{+}}={-\bf \hat{i}})}{d\hat{s}}\Big]}.
\eea
Similarly, the double spin asymmetries ${\cal P}_{ij}$ can be
obtained:
\beq {\cal P}_{ij}
=\frac{\Big[\frac{d\Gamma({\bf s^{+}}={\bf \hat{i}},{\bf
      s^{-}}={\bf \hat{j}})}{d\hat{s}}-\frac{d\Gamma({\bf s^{+}}={\bf
      \hat{i}},{\bf s^{-}}={-\bf \hat{j}})}{d\hat{s}}\Big]
  -\Big[\frac{d\Gamma({\bf s^{+}}={-\bf \hat{i}},{\bf s^{-}}={\bf
      \hat{j}})}{d\hat{s}}-\frac{d\Gamma({\bf s^{+}}={-\bf
      \hat{i}},{\bf s^{-}}={-\bf \hat{j}})}{d\hat{s}}\Big]}
{\Big[\frac{d\Gamma({\bf s^{+}}={\bf \hat{i}},{\bf s^{-}}={\bf
      \hat{j}})}{d\hat{s}}+\frac{d\Gamma({\bf s^{+}}={\bf
      \hat{i}},{\bf s^{-}}={-\bf \hat{j}})}{d\hat{s}}\Big]
  +\Big[\frac{d\Gamma({\bf s^{+}}={-\bf \hat{i}},{\bf s^{-}}={\bf
      \hat{j}})}{d\hat{s}}+\frac{d\Gamma({\bf s^{+}}={-\bf
      \hat{i}},{\bf s^{-}}={-\bf \hat{j}})}{d\hat{s}}\Big]},
\eeq
where ${\hat i}$ and ${\hat j}$ are unit vectors along the $i$ and $j$
directions. Note that both ${\cal P}_{i}^{\pm}$ and ${\cal P}_{ij} $
depend also on $\hat{s}$.  However, the explicit dependence on
$\hat{s}$ has been suppressed for simplicity of notation.

Before presenting explicit expressions for these quantities, it is
useful to make the following remark. With our choice of 4-momenta
[Eq.~(\ref{frame})], the decay takes place in the $yz$ plane.
Therefore, the only vectors which can have ${\hat x}$ components are
the spins ${\bf s^+}$ and ${\bf s^-}$. This implies that the only
scalar product which involves $\hat x$ components is the dot product
of two spins. Thus, any term that has only one component of spin along
$\hat x$ (i.e.\ ${\cal P}_x$, ${\cal P}_{xy}$ and ${\cal P}_{xz}$)
must come from a triple-product correlation. This holds even in the
presence of new physics. It is therefore these quantities which probe
the imaginary parts of the products of Wilson coefficients.

The ${\cal P}$'s take the form
\bea
{\cal P}^{+}_x &=&\frac{-3\,\pi}{2\,\sqrt{\hat{s}}\Delta } \,\Big(
    2\,{{\rm Im}(C_7^{eff}C_{10}^*)} + {{\rm
    Im}(C_9^{eff}C_{10}^*)}\,\hat{s}\Big)\,{\hat{m}_\tau}\, \sqrt{1 -
    \frac{4\,{{\hat{m}_\tau}}^2}{\hat{s}}}
\label{firstP}
\\
{\cal P}^{+}_y &=& \frac{3\,\pi}{2\,\sqrt{\hat{s}}\Delta }
   \,\Big(\frac{ 4\,|C_7^{eff}|^2}{\hat{s}}+2\,{{\rm Re}(C_7^{eff}
   C_{10}^*)} +4\,{{\rm Re}(C_7^{eff} C_9^{eff^*})} \nn\\ &&
   \hskip1.7truein + ~{{\rm Re}(C_9^{eff} C_{10}^*)}
   +|C_9^{eff}|^2\,\hat{s}\Big)\, {\hat{m}_\tau} \\
{\cal P}^{+}_z &=&\frac{2}{\Delta}\,\Big( 6\,{{\rm Re}(C_7^{eff}
    C_{10}^*)} + {{\rm Re}(C_9^{eff} C_{10}^*)}\,( 1 +
    2\,\hat{s})\Big)\, {\sqrt{1 -
    \frac{4\,\hat{m}_\tau^2}{\hat{s}}}}\\
{\cal P}^{-}_x &=&{\cal P}^{+}_x \\
{\cal P}^{-}_y &=& \frac{3\,\pi}{2\,\sqrt{\hat{s}}\Delta }
   \,\Big(\frac{ 4\,|C_7^{eff}|^2}{\hat{s}}-2\,{{\rm
   Re}(C_7^{eff}C_{10}^*)} +4\,{{\rm Re}(C_7^{eff} C_9^{eff^*})} \nn\\
   && \hskip1.7truein -~{{\rm Re}(C_9^{eff} C_{10}^*)}
   +|C_9^{eff}|^2\,\hat{s}\Big)\, {\hat{m}_\tau} \\
{\cal P}^{-}_z &=&{\cal P}^{+}_z \\
{\cal P}_{xx}&=&\frac{1}{\Delta}\Bigg(24\,{{\rm Re}(C_7^{eff}
C_9^{eff^*})}\,\frac{\hat{m}_\tau^2}{\hat{s}}+
4\,|C_7^{eff}|^2\,\frac{ \Big( ( -1 + {\hat{s}}) \, {{\hat{s}}} + 2\,(
2 + {\hat{s}} ) \,{{\hat{m}_\tau}}^2 \Big) }{{{\hat{s}}}^2} \nn \\ &&
\quad + (|C_9^{eff}|^2-|C_{10}|^2)\,\frac{\Big( (1
-{{\hat{s}}}){\hat{s}} + 2\,(1+2\,{\hat{s}}){{\hat{m}_\tau}}^2 \Big)
}{{{\hat{s}}}} \Bigg) \\
{\cal P}_{yx}&=&\frac{-2}{\Delta}\,{{\rm Im}(C_9^{eff} C_{10}^*)}\,( 1
  - \hat{s}) \, \sqrt{1 - \frac{4\,\hat{m}_\tau^2}{\hat{s}}} \\
{\cal P}_{zx}&=&\frac{-3\,\pi} {2\,\sqrt{\hat{s}}\Delta}\,\Big(
    2\,{{\rm Im}(C_7^{eff} C_{10}^*)} + {{\rm Im}(C_9^{eff} C_{10}^*)}
    \Big) \,{\hat{m}_\tau} \\
{\cal P}_{xy}&=&{\cal P}_{yx}\\
{\cal P}_{yy}&=&\frac{1}{\Delta}\Bigg( \frac{24\,{{\rm Re}(C_7^{eff}
C_9^{eff^*})}\,\hat{m}_\tau^2}{\hat{s}} -
4\,(|C_9^{eff}|^2+|C_{10}|^2)\,\frac{ ( 1 - \hat{s})\,\hat{m}_\tau^2}
{\hat{s}} \\&& + (|C_9^{eff}|^2-|C_{10}|^2)\,( (-1 + \hat{s})+
\frac{6\,{{\hat{m}_\tau}}^2}{\hat{s}})\nn \\&& \quad +
\frac{4\,|C_7^{eff}|^2\, \Big( ( 1 - \hat{s}) \, \hat{s} + 2\,( 2 +
\hat{s} ) \,\hat{m}_\tau^2 \Big) } {{\hat{s}}^2} \Bigg) \\
{\cal P}_{zy}&=&\frac{3\,\pi }{2\,\sqrt{\hat{s}}\Delta}\,\Big(2\,
    {{\rm Re}(C_7^{eff} C_{10}^*)}-|C_{10}|^2+{{\rm Re}(C_9^{eff}
    C_{10}^*)}\, \hat{s}\Big)\,{\hat{m}_\tau}\,\sqrt{1 -
    \frac{4\,\hat{m}_\tau^2}{\hat{s}}} \\
{\cal P}_{xz}&=&-{\cal P}_{zx}\\
{\cal P}_{yz}&=&\frac{3\,\pi}{2\,\sqrt{\hat{s}}\Delta} \Big(2\, {{\rm
   Re}(C_7^{eff} C_{10}^*)}+|C_{10}|^2+{{\rm Re}(C_9^{eff}
   C_{10}^*)}\, \hat{s}\Big)\,{\hat{m}_\tau}\,\sqrt{1 -
   \frac{4\,\hat{m}_\tau^2}{\hat{s}}} \\
{\cal P}_{zz}&=&\frac{1}{\Delta} \Bigg( 12\,{{\rm Re}(C_7^{eff}
C_9^{eff^*})}\, \Big( 1 - \frac{2\,{{\hat{m}_\tau}}^2}{\hat{s}} \Big)
+ \frac{4\,|C_7^{eff}|^2 \, ( 2 + \hat{s}) \, \Big( 1 -
\frac{2\,\hat{m}_\tau^2}{\hat{s}} \Big) } {\hat{s}} \\ &&+
(|C_9^{eff}|^2+|C_{10}|^2)\, \Big( 1 + 2\,\hat{s} - \frac{6\,( 1 +
\hat{s}) \,\hat{m}_\tau^2} {\hat{s}} \Big) \nn \\ &&+
\frac{2\,(|C_9^{eff}|^2-|C_{10}|^2)\,(2+\hat{s})\,\hat{m}_\tau^2}
{\hat{s}} \Bigg) ~. 
\label{lastP}
\eea

\begin{figure}[htb]
\begin{center}
  \includegraphics*[scale=0.8]{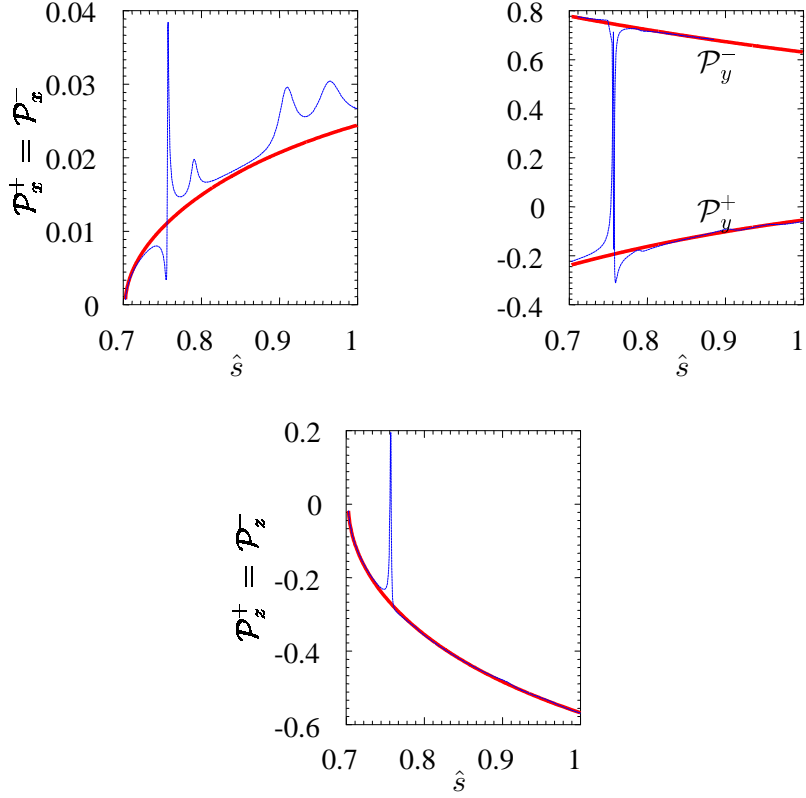}
\end{center}
\caption{The polarization asymmetries for the $\tau^-$ and $\tau^+$,
  as functions of ${\hat s}$, the invariant mass of the $\tau$ pair,
  without (thick lines) and with (thin lines) the long-distance
  resonance contributions.} 
\label{fig1}
\end{figure}

The coefficient ${\cal P}_z^-$ was computed in Ref.~\cite{Hewett},
${\cal P}_x^-$, ${\cal P}_y^-$ and ${\cal P}_z^-$ were obtained in
Ref.~\cite{KS}, and ${\cal P}_x^+$, ${\cal P}_y^+$ and ${\cal P}_z^+$
were calculated in Ref.~\cite{FKY}. (Note: while we agree with the
calculations of Refs.~\cite{Hewett,KS}, we disagree with
Ref.~\cite{FKY} about the expression for ${\cal P}_y^\pm$ [the
equation following their Eq.~(24)].) We plot the functions ${\cal
P}_x^-$, ${\cal P}_y^-$ and ${\cal P}_z^-$ as functions of $\hat{s}$
in Fig.~\ref{fig1}. For the purpose of numerical computations, we
follow the prescription of Ref.~\cite{KS} and include long-distance
effects in $C_9^{eff}$ associated with real $c\bar{c}$ resonances in
the intermediate states.  We take the phenomenological parameter
$\kappa_V$ multiplying the Breit-Wigner function in Ref.~\cite{KS} to
be unity.

\begin{figure}[!ht]
\begin{center}
  \includegraphics*[scale=0.8]{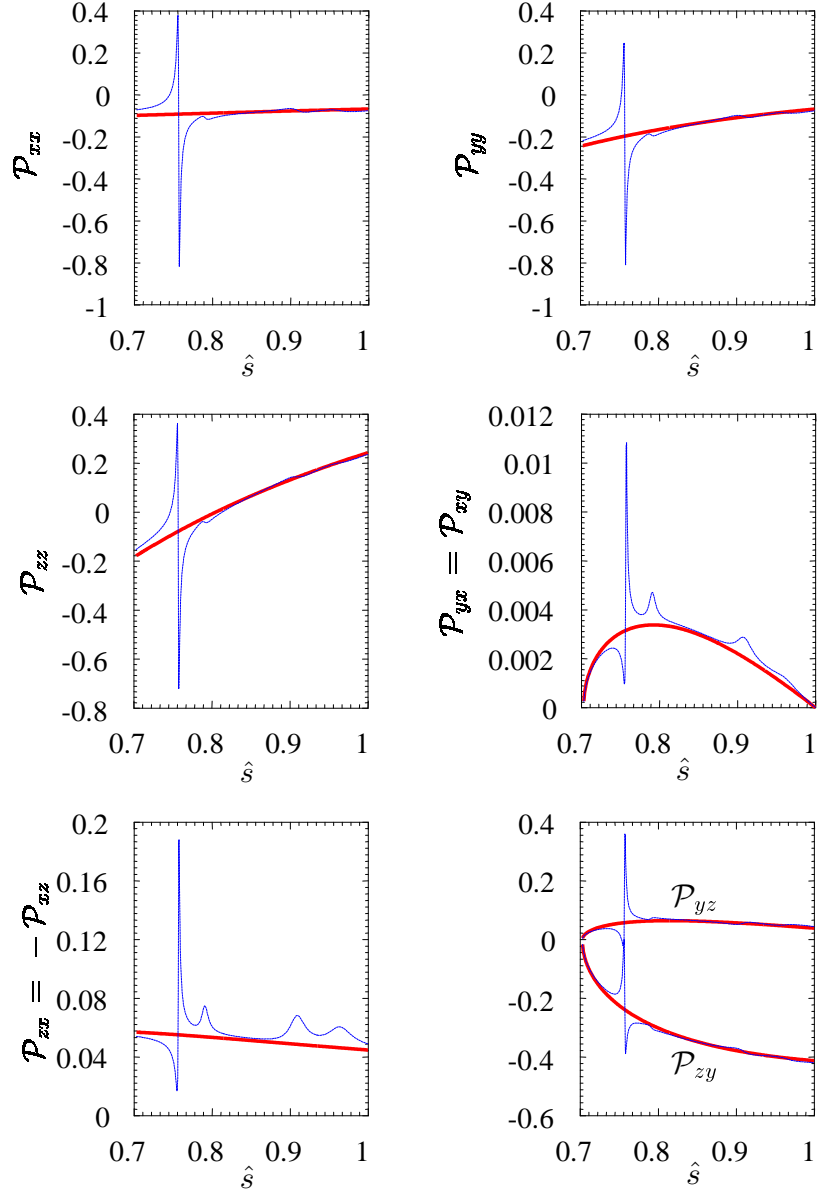}
\end{center}
\caption{The double-spin polarization asymmetries, as functions of
  ${\hat s}$, the invariant mass of the $\tau$ pair, without (thick
  lines) and with (thin lines) the long-distance resonance
  contributions.}
\label{fig2}
\end{figure}

Note that our ${\cal P}_z^-$ is the same as the longitudinal
polarization asymmetry of the $\tau^-$, $P_L^-$ of
Refs.~\cite{KS,FKY}. However, ${\cal P}_z^+=-P_L^+$, since the
$\tau^+$ moves along the $-\hat{z}$ axis. Similarly, ${\cal
P}_x^-=P_N^-$ and ${\cal P}_x^+=-P_N^+$. (Note: the distribution of
our $P_N$ differs from that of Ref.~\cite{KS}, resulting in a somewhat
smaller value of $\langle P_N \rangle_\tau$.) The transverse direction
defined in these references lies along the negative ${\hat
y}$-direction for both $\tau^-$ and $\tau^+$, so that ${\cal
P}_y^-=-P_T^-$ and ${\cal P}_y^+=-P_T^+$. The double-spin polarization
asymmetries are shown in Fig.~\ref{fig2}.

Previously, we noted that it is very likely that only asymmetries
larger than 10\% will be measurable. One characterization of the data
is to calculate the average values of the above
asymmetries\footnote{It is also possible that the average value of an
asymmetry is small, but that large values of the asymmetry are still
possible for certain values of ${\hat s}$. For example, see ${\cal
P}_{zz}$.}. These are defined as
\beq
\langle{\cal P}\rangle \equiv \dsp \frac{\dsp
  \int^1_{4\,\hat{m}_\tau^2}{\cal
  P}\dsp\frac{d\Gamma}{d\hat{s}}d\hat{s}}{\dsp
  \int^1_{4\,\hat{m}_\tau^2} \dsp
  \frac{d\Gamma}{d\hat{s}}d\hat{s}}~.
\label{eq:average}
\eeq
In Table~\ref{tab1} we list the average values of all polarization
asymmetries. From this Table, we see that only ${\cal P}_y^\pm$,
${\cal P}_z$, ${\cal P}_{yy}$, and ${\cal P}_{zy}$ can be considered
sizeable. (Note that here, ${\cal P}_{z}^+ = {\cal P}_{z}^- \equiv
{\cal P}_{z}$.)

\begin{table}[!tbh]
\begin{center}
\begin{tabular}{|c|c|}
\hline
\hline
~~~~~~~~~~~~~~~~~~~~~~~~~~~~~~~~&~~~~~~~~~~~~~~~~~~~~~~~~~~~~~~~~~~~\\
$\langle{\cal P}_x^-\rangle=\langle{\cal P}_x^+\rangle$ & $1.413\times
10^{-2}$ \\[2ex]
$\langle{\cal P}_y^-\rangle$ & 0.723 \\[2ex]
$\langle{\cal P}_z^-\rangle=\langle{\cal P}_z^+\rangle$ & $-$0.336 \\
[2ex]
$\langle{\cal P}_y^+\rangle$ & $-$0.164 \\[2ex]
$\langle{\cal P}_{xx}\rangle$ & $-8.658\times 10^{-2}$ \\[1ex]
$\langle{\cal P}_{yx}\rangle=\langle{\cal P}_{xy}\rangle$ &
$2.868\times 10^{-3}$\\ [2ex]
$\langle{\cal P}_{zx}\rangle=-\langle{\cal P}_{xz}\rangle$ &
$5.322\times 10^{-2}$ \\ [2ex]
$\langle{\cal P}_{yy}\rangle$ & $-$0.168 \\ [2ex]
$\langle{\cal P}_{zy}\rangle$ & $-$0.281 \\ [2ex]
$\langle{\cal P}_{yz}\rangle$ & $5.717\times 10^{-2}$ \\ [2ex]
$\langle{\cal P}_{zz}\rangle$ & $-1.1254\times 10^{-2}$ \\ [2ex]
\hline
\hline
\end{tabular}
\caption{Numerical values of the various averaged spin-polarization
  asymmetries without including the long-distance resonance
  contributions. We use $m_b=4.24$ GeV \cite{luke}. The corresponding
  branching ratio is $BR(B\to X_s\tau^+\tau^-)=1.192\times 10^{-7}$.}
\label{tab1}
\end{center}
\end{table}

In Eqs.~(\ref{firstP})--(\ref{lastP}), there are certain relations
between the ${\cal P}$'s when the spins ${\bf s^{+}}$ and ${\bf
s^{-}}$ are interchanged. Some of these relations are equalities,
e.g.\ ${\cal P}_x^- = {\cal P}_x^+$, ${\cal P}_{xz} = {\cal P}_{zx}$,
etc. For other pairs of ${\cal P}_i$'s, the expressions are similar,
but only some of the terms change sign (e.g.\ ${\cal P}_y^+$
vs. ${\cal P}_y^-$). As we describe below, it is possible to
understand these relations by considering also the conjugate process
${\bar b} \to {\bar s} \tau^- \tau^+$.

The processes $b \to s \tau^+ \tau^-$ and ${\bar b} \to {\bar s}
\tau^- \tau^+$ are related by CPT as follows \cite{CPTref}:
\bea
b(p_b) & \to & s(p_s)\,\tau^+(p_+,{\bf s_+})\,\tau^-(p_-,{\bf s_-}) ~,
\nn\\
\bar{b}(p_b) & \to & \bar{s}(p_s)\,\tau^-(p_+,-{\bf
s_+})\,\tau^+(p_-,-{\bf s_-}) ~.
\label{CPTrels}
\eea
In the absence of CP violation, observables which are P-odd must
vanish in the (C-even) untagged sample. Consider first the terms
involving triple-product (TP) correlations. While all triple products
are T-odd, they can be either P-even or P-odd. Triple products
involving two spins are necessarily P-odd and, in the absence of CP
violation, C-odd.  Because of this, in the SM, these triple product
must vanish in the untagged sample. Thus, we have ${\rm
TP}^{P-odd}_{\bar b} = -{\rm TP}^{P-odd}_b$. This relation can be
violated in the presence of CP-violating new physics. On the other
hand, triple products involving one spin are P-even and C-even, so
that ${\rm TP}^{P-even}_{\bar b} = +{\rm TP}^{P-even}_b$, in the
absence of CP violation. Thus, these triple products can survive in
the untagged sample due to the presence of the strong phases which can
fake CP-violating effects.

We now apply these observations to ${\cal P}_x^+$ and ${\cal P}_x^-$,
which involve a single spin. As noted earlier, terms with a single
spin along $\hat x$ must come only from a triple-product correlation.
The general triple-product term giving these quantities can be written
as $\epsilon_{\alpha\beta\mu\rho} \, p_b^\alpha p_s^\beta (a\,p_+^\mu
s_+^\rho +b\, p_-^\mu s_-^\rho)$, where $a$ and $b$ are arbitrary
coefficients. For the conjugate process [Eq.~(\ref{CPTrels})], the
corresponding term is $-\epsilon_{\alpha\beta\mu\rho} \, p_b^\alpha
p_+^\beta (a\,p_-^\mu s_-^\rho +b\,p_+^\mu s_+^\rho)$. Since ${\rm
TP}^{P-even}_{\bar b} = +{\rm TP}^{P-even}_b$, this implies that
$a=-b$ (in the absence of CP violation). Using the 4-vectors of
Eq.~(\ref{frame}), it is then straightforward to show that this
results in ${\cal P}_x^+ = +{\cal P}_x^-$. Note that this will hold
even in the presence of CP-conserving New Physics.

Similarly, the two-spin triple products, which contribute to the pairs
$\{{\cal P}_{yx},{\cal P}_{xy}\}$ and $\{{\cal P}_{zx},{\cal
P}_{xz}\}$, are proportional to $\epsilon_{\alpha\beta\mu\rho} \,
p_s^\alpha p_b^\beta s_-^\mu s_+^\rho$. In the absence of CP
violation, the CP-odd combination of ${\cal P}_{yx}$ and ${\cal
P}_{xy}$ (and of ${\cal P}_{zx}$ and ${\cal P}_{xz}$) will vanish in
an untagged sample.  Again, a simple calculation then shows that this
implies that ${\cal P}_{yx}= +{\cal P}_{xy}$ and ${\cal P}_{zx}=
-{\cal P}_{xz}$.

For the other terms that do not contain triple products, and are hence
always T-even, one can understand the relationship between the ${\cal
P}$'s in a similar fashion. For example, consider ${\cal P}_y^+$ and
${\cal P}_y^-$. Since only dot products of various momenta and one
spin are involved, the coefficients of both terms $|C_7^{eff}|^2$
[Eq.~(\ref{T7squared})] and ${\rm Re}(C_7^{eff} C_{10}^*)$
[Eq.~(\ref{T7T10})] are T-even and P-odd. However, the $|C_7^{eff}|^2$
term ``$p_s\cdot(s^- + s^+)$'' switches sign under CPT for the
conjugate process, while the ${\rm Re}(C_7^{eff} C_{10}^*)$ term
``$p_s\cdot(s^+ - s^-)$'' has the same sign for the conjugate process.
Since these terms are P-odd and C-odd (in the absence of CP
violation), they must vanish in an untagged sample. This explains the
relative sign difference between the $|C_7^{eff}|^2$ and ${\rm
Re}(C_7^{eff} C_{10}^*)$ terms in ${\cal P}_y^+$ and ${\cal
P}_y^-$. This argument may be extended to all terms contributing to
various ${\cal P}_i$'s. In particular, in the SM, ${\cal P}_z^+ =
+{\cal P}_z^-$. On the other hand, in presence of New Physics, while
the additional terms must still be T-even and P-odd, they could be
even or odd under CPT, implying that the relation between ${\cal
P}_z^+$ and ${\cal P}_z^-$ could differ.
 
Of course, the above discussion assumes that there is no CP violation
in $b \to s \tau^+ \tau^-$, which is the case in the SM, to a good
approximation. On the other hand, if new CP-violating physics
contributes to this decay, this gives us several clear tests for its
presence. For example, any violation of the relation ${\cal P}_x^+ =
{\cal P}_x^-$ (or $P_L^- + P_L^+=0$) is a smoking-gun signal of such
new physics.

\section{Forward-Backward Asymmetries}

One observable which does not depend on the polarization of the
final-state leptons is the forward-backward (FB) asymmetry. In the
frame of reference described in Eq.~(\ref{frame}), the
forward-backward asymmetry is given by
\bea
A_{FB}(\hat{s}) &=& \dsp\frac{\dsp\int_{0}^1 \frac
{d^2\Gamma}{d\hat{s}~d\cos\theta}d\cos\theta- \int_{-1}^{0}\frac
{d^2\Gamma}{d\hat{s}~d\cos\theta}d\cos\theta} {\dsp\int_{0}^1 \frac
{d^2\Gamma}{d\hat{s}~d\cos\theta}d\cos\theta+ \int_{-1}^{0}\frac
{d^2\Gamma}{d\hat{s}~d\cos\theta}d\cos\theta} \nn\\
&=& \frac{3}{\Delta}\,\Big(2\,{{\rm Re}(C_7^{eff}
C_{10}^*)}\,+\hat{s}\,{{\rm Re}(C_9^{eff} C_{10}^*)}\Big)\,\sqrt{1 -
\frac{4\,\hat{m}_\tau^2}{\hat{s}}} ~.
\label{FBasym}
\eea
This agrees with the result of Ref.~\cite{KS} (and that of
Ref.~\cite{Ali} when $m_\tau$ is neglected). Note that the FB
asymmetry is of opposite sign for the CP-conjugate process ${\bar b}
\to {\bar s} \tau^+ \tau^-$, so that $A_{FB}^b + A_{FB}^{\bar b} = 0$.
Thus, in order to measure the unpolarized FB asymmetry, it will be
necessary to tag the flavor of the decaying $b$-quark.

If the polarization of the final-state leptons can be measured, then,
in addition to the polarization asymmetries discussed in the previous
section, one can also extract forward-backward asymmetries of the
polarized leptons. While the unpolarized FB asymmetry of
Eq.~(\ref{FBasym}) requires $b$-tagging, some of the polarized FB
asymmetries are non-vanishing even in an untagged sample.

We can extract the forward-backward asymmetries corresponding to
various polarization components of the $\tau^-$ and/or $\tau^+$ spin
by writing:
\bea
  A_{FB}({\bf s^{+}},{\bf s^{-}},\hat{s})&=& A_{FB}(\hat{s})+ \Big[
{\cal A}^{-}_x s_x^- +{\cal A}^{-}_y s_y^- +{\cal A}^{-}_z s_z^-
+{\cal A}^{+}_x s_x^+ +{\cal A}^{+}_y s_y^+ +{\cal A}^{+}_z s_z^+ \nn
\\
&& \hskip1truein +~{\cal A}_{xx} s_x^+ s_x^- +{\cal A}_{xy} s_x^+
                s_y^- +{\cal A}_{xz} s_x^+ s_z^- \nn\\
&& \hskip1truein +~{\cal A}_{yx} s_y^+ s_x^- +{\cal A}_{yy} s_y^+
                s_y^- +{\cal A}_{yz} s_y^+ s_z^- \nn \\
&& \hskip1truein +~{\cal A}_{zx} s_z^+ s_x^- +{\cal A}_{zy} s_z^+
                s_y^- +{\cal A}_{zz} s_z^+ s_z^- \Big] ~.
\eea
The various polarized forward-backward asymmetries are then evaluated
to be
\bea
{\cal A}^{+}_x &=& 0 \\
{\cal A}^{+}_y &=& \frac{2}{\Delta}\,{{\rm Re}(C_9^{eff}
     C_{10}^*)}\,\frac{(1-\hat{s})\,
     \hat{m}_\tau}{\sqrt{\hat{s}}}\,\sqrt{1 -
     \frac{4\,\hat{m}_\tau^2}{\hat{s}}} \\
{\cal A}^{+}_z &=&\frac{1}{\Delta}\Bigg( 6\,{{\rm Re}(C_7^{eff}
     C_9^{eff^*})} - \frac{6\,|C_7^{eff}|^2}{\hat{s}} -
     3\,(\,|C_9^{eff}|^2-|C_{10}|^2)\,\hat{m}_\tau^2 \nn\\ && -
     12\,{{\rm Re}(C_7^{eff} C_{10}^*)}\,\frac{\hat{m}_\tau^2}
     {\hat{s}} - ~6\,{{\rm Re}(C_9^{eff}
     C_{10}^*)}\,\frac{\hat{m}_\tau^2} {\hat{s}} \nn \\ &&
     -~\frac{3}{2}\,(\,|C_9^{eff}|^2+|C_{10}|^2)\,\hat{s}\, ( 1 -
     \frac{2\,\hat{m}_\tau^2}{\hat{s}})\Bigg) \\
{\cal A}^{-}_x &=& 0\\
{\cal A}^{-}_y &=& {\cal A}^{+}_y \\
{\cal A}^{-}_z &=&\frac{1}{\Delta}\Bigg( -6\,{{\rm Re}(C_7^{eff}
     C_9^{eff^*})} - \frac{6\,|C_7^{eff}|^2}{\hat{s}} -
     3\,(\,|C_9^{eff}|^2-|C_{10}|^2)\,\hat{m}_\tau^2 \nn\\ &&
     +~12\,{{\rm Re}(C_7^{eff} C_{10}^*)}\,\frac{\hat{m}_\tau^2}
     {\hat{s}} +~6\,{{\rm Re}(C_9^{eff}
     C_{10}^*)}\,\frac{\hat{m}_\tau^2} {\hat{s}} \nn \\ &&
     -~\frac{3}{2}\,(\,|C_9^{eff}|^2+|C_{10}|^2)\,\hat{s}\, ( 1 -
     \frac{2\,\hat{m}_\tau^2}{\hat{s}})\Bigg) \\
{\cal A}_{xx} &=& 0\\
{\cal A}_{xy} &=& \frac{-6}{\Delta}\,( 2\,{{\rm
Im}(C_7^{eff}\,C_{10}^*)} + {{\rm
Im}(C_9^{eff}\,C_{10}^*)})\,\frac{{{\hat{m}_\tau}}^2}{\hat{s}} \\
{\cal A}_{xz} &=& \frac{2}{\Delta}\,{{\rm
     Im}(C_9^{eff}C_{10}^*)}\,\frac{(1-\hat{s})\,{\hat{m}_\tau}
     }{\sqrt{\hat{s}}}\,\sqrt{1 - \frac{4\,\hat{m}_\tau^2}{\hat{s}}}
     \\
{\cal A}_{yx} &=& -{\cal A}_{xy}\\
{\cal A}_{yy} &=& 0\\
{\cal A}_{yz}\! &=& \!\Big( 2
      |C_9^{eff}|^2-\frac{8\,|C_7^{eff}|^2}{\hat{s}} \Big) \, \frac{(
      1 - \hat{s})\,\hat{m}_\tau}{\Delta\sqrt{\hat{s}}} \\
{\cal A}_{zx} &=& {\cal A}_{xz}\\
{\cal A}_{zy} &=& {\cal A}_{yz} \\
{\cal A}_{zz} &=& \frac{-3}{\Delta}\,( 2\,{{\rm
     Re}(C_7^{eff}\,C_{10}^*)} + {{\rm
     Re}(C_9^{eff}\,C_{10}^*)}\,\hat{s}) \, {\sqrt{1 -
     \frac{4\,{{\hat{m}_\tau}}^2}{\hat{s}}}} ~.
\eea
Note that, in the SM, it turns out that $A_{FB} = -{\cal A}_{zz}$.

\begin{figure}[!htb]
\begin{center}
\includegraphics*[scale=0.8]{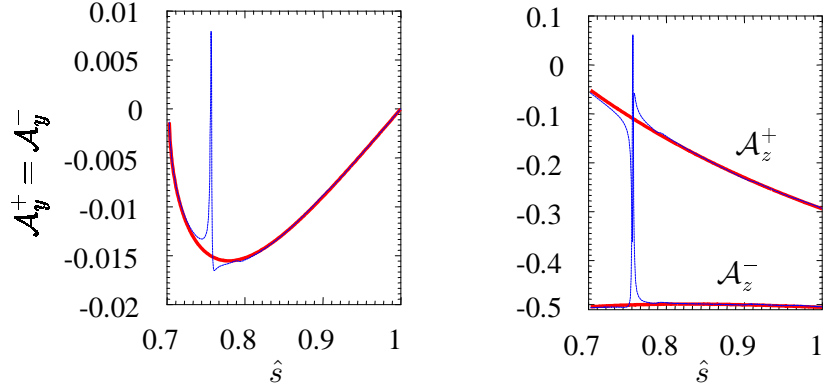}
\end{center}
\caption{Forward-backward asymmetries of the $\tau^-$ and $\tau^+$, as
   functions of ${\hat s}$, the invariant mass of the $\tau$ pair,
   without (thick lines) and with (thin lines) the long-distance
   resonance contributions.}
\label{fig3}
\end{figure}

\begin{figure}[!ht]
\begin{center}
\includegraphics*[scale=0.8]{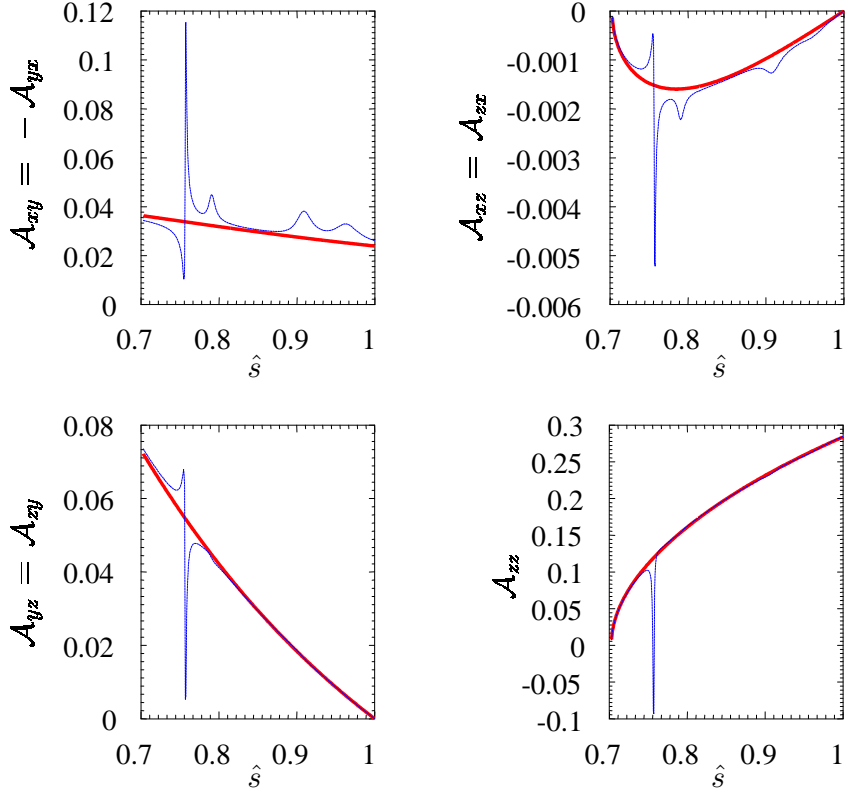}
\end{center}
\caption{Doubly-polarized forward-backward asymmetries, as functions
    of ${\hat s}$, the invariant mass of the $\tau$ pair, without
    (thick lines) and with (thin lines) the long-distance
    resonance contributions.}
\label{fig4}
\end{figure}

The nonzero single-spin forward-backward asymmetries are depicted in
Fig.~\ref{fig3} as functions of $\hat{s}$, while those with both spins
polarized are shown in Fig.~\ref{fig4}. Interestingly, some of the
forward-backward asymmetries are identically zero within the SM:
${\cal A}_{x}^+$, ${\cal A}_{x}^-$, ${\cal A}_{xx}$ and ${\cal
A}_{yy}$. Nonvanishing values of these asymmetries would be clear
signals of NP. Also, as was discussed in the case of the polarization
asymmetries, the discrete transformation properties of the operators
can once again be used to understand the relations between pairs of
forward-backward asymmetries in which the spins ${\bf s^{+}}$ and
${\bf s^{-}}$ are interchanged.

The average values of the forward-backward asymmetries are defined
similarly to Eq.~(\ref{eq:average}) and are listed in
Table~\ref{tab2}. From this table, we see that only three asymmetries
are expected to be larger than 10\% in the SM: ${\cal A}_z^\pm$ and
${\cal A}_{zz}$.

\begin{table}[htb]
\begin{center}
\begin{tabular}{|c|c|}
\hline
\hline
~~~~~~~~~~~~~~~~~~~~~~~~~~~~~~~~&~~~~~~~~~~~~~~~~~~~~~~~~~~~~~~~~~~~\\
$ \langle{\cal A}_y^+\rangle=\langle{\cal A}_y^-\rangle$ &
$-1.302\times 10^{-2}$ \\ [2ex]
$ \langle{\cal A}_z^+\rangle$ & $-$0.148 \\ [2ex]
$ \langle{\cal A}_z^-\rangle$ & $-$0.490 \\ [2ex]
$ \langle{\cal A}_{xy}\rangle=-\langle{\cal A}_{yx}\rangle$ &
$3.184\times 10^{-2}$ \\ [2ex]
$ \langle{\cal A}_{xz}\rangle=\langle{\cal A}_{zx}\rangle$ &
$-1.347\times 10^{-3}$ \\ [2ex]
$ \langle{\cal A}_{yz}\rangle=\langle{\cal A}_{zy}\rangle$ &
$4.298\times 10^{-2}$ \\ [2ex]
$ \langle{\cal A}_{zz}\rangle$ & 0.154 \\ [2ex]
\hline 
\hline
\end{tabular}
\caption{Numerical values of the various average polarized
  forward-backward asymmetries without including the long distance
  resonance contributions. We use $m_b=4.24$ GeV \cite{luke}. The
  corresponding average unpolarized forward-backward asymmetry is
  $\langle A_{FB} \rangle =-0.154$.}
\label{tab2}
\end{center}
\end{table}

\section{Discussion}

In the previous sections, we have discussed the polarization and
forward-backward asymmetries which can be obtained when the spins of
the $\tau^+$ and/or $\tau^-$ are measured. Here we consider what can
be learned from these measurements in a variety of scenarios.

First, suppose that the statistics are such that only a single
polarization can be measured (say that of the $\tau^-$), and that no
tagging is possible. In this case only the P-even observables survive:
${\cal P}_x^+ + {\cal P}_x^-$, ${\cal A}_y^+ + {\cal A}_y^-$ and
${\cal A}_z^+ + {\cal A}_z^-$. Of these asymmetries only ${\cal A}_z^+
+ {\cal A}_z^-$ is measurable within the SM. Along with the
differential decay rate, this therefore gives only two observables,
which is not enough to test the SM.

On the other hand, if the polarizations of both $\tau^+$ and $\tau^-$
can be measured, still without tagging, then one adds another six
observables: ${\cal P}_{xx}$, ${\cal P}_{yy}$, ${\cal P}_{zy} + {\cal
P}_{yz}$, ${\cal P}_{zz}$, ${\cal A}_{xy} + {\cal A}_{yx}$, and ${\cal
A}_{xz} +{\cal A}_{zx}$. Of these only three --- ${\cal P}_{yy}$,
${\cal P}_{zy} + {\cal P}_{yz}$ and ${\cal P}_{zz}$ --- are expected
to be sizeable in the SM, the last one being measurable only as a
distribution in $\hat{s}$ (see Fig.~\ref{fig2}). We therefore have
just enough measurements to determine the five unknowns $C_7^{eff}$,
$C_{10}$, ${\rm Re}(C_9^{eff})$, ${\rm Im}(C_9^{eff})$, and $m_b$.
However, there are not enough measurements to provide an internal
crosscheck of the predictions of the SM.

Now suppose that it is possible to tag the flavor of the decaying
$b$-quark. If only a single $\tau$-spin measurement is performed then,
out of a total of thirteen possible asymmetries, only six are sizeable
within the SM: ${\cal A}_{FB}$, ${\cal P}_{y}^\pm$, ${\cal P}_{z}$ and
${\cal A}_{z}^\pm$. (Recall that ${\cal P}_{z}^+ = {\cal P}_{z}^-
\equiv {\cal P}_{z}$.)

In the best-case scenario, it will be possible to both tag the flavor
of the decaying $b$, and to measure the polarizations of both
final-state $\tau$ leptons. In this case, one in principle has 31
asymmetries. However, within the SM only nine of these are accessible:
${\cal A}_{FB}$, ${\cal P}_{y}^\pm$, ${\cal P}_{z}$, ${\cal
A}_{z}^\pm$, ${\cal P}_{yy}$, ${\cal P}_{zy}$ and ${\cal A}_{zz}$.
Even so, if these asymmetries could be measured, this would allow us
to greatly overconstrain the SM. Ideally, we will find evidence for
new physics, but if not, these will provide precision determinations
of both $m_b$ and the Wilson coefficients describing the decay $b\to s
\tau^+ \tau^-$.

In Table~\ref{tab3} we summarize the number of possible observables,
including the differential cross-section, in the various scenarios
discussed above.

\begin{table}[!htb]
\begin{center}
\begin{tabular}{|p{2.5cm}|c|p{4.0cm}|p{4.0cm}|}
\hline
\hline & & &\\
& &Untagged Sample & Tagged Sample\\[2ex]
\hline & & &\\
Only one of $\tau^+$ or $\tau^-$ spin measured&
  &$\frac{d\Gamma}{d\hat{s}}$, ${\cal P}_x^{(\pm)}$, ${\cal
  A}_y^{(\pm)}$, ${\cal A}_z^{(\pm)}$[4] &$\frac{d\Gamma}{d\hat{s}}$,
  ${\cal A}_{FB}$, ${\cal P}_x^\pm$, ${\cal P}_y^\pm$, ${\cal
  P}_z^\pm$, ${\cal A}_x^\pm$, ${\cal A}_y^\pm$, ${\cal A}_z^\pm$
  [14]\\[2ex]
\cline{2-4}& && \\ &SM &$\frac{d\Gamma}{d\hat{s}}$, ${\cal
 A}_{z}^{(\pm)}$ [2]&$\frac{d\Gamma}{d\hat{s}}$, ${\cal A}_{FB}$,
 ${\cal P}_{y}^\pm$, ${\cal P}_{z}$, ${\cal A}_{z}^\pm$ [7]\\[2ex]
\hline & & &\\
Both $\tau^+$ and $\tau^-$ spins measured&& $+$~${\cal P}_{xx}$,
  ${\cal P}_{yy}$, ${\cal P}_{(zy)}$, ${\cal P}_{zz}$, ${\cal
  A}_{(xy)}$, ${\cal A}_{(xz)}$ [10]& All [32]\\[2ex]
\cline{2-4} &&&\\
  &SM&$+$~${\cal P}_{yy}$, ${\cal P}_{(zy)}$, ${\cal P}_{zz}$
  [5]&$+$~${\cal P}_{yy}$, ${\cal P}_{zy}$,
${\cal P}_{zz}$ [10] \\[2ex]
\hline
\hline
\end{tabular}
\caption{The number of observables in various scenarios of initial
  $b$-flavor tagging and $\tau$-spin measurements. The columns
  represent the cases with untagged and tagged samples, while the rows
  are for the scenarios in which only one of the $\tau^+$ or $\tau^-$
  spin is measured, or when both $\tau^+$ and $\tau^-$ spins are
  measured. Of the possible observables, those that are sizeable in
  the SM are listed separately. In the case in which both spins are
  measured, only the additional observables (indicated by $+$) are
  listed. The number in the square brackets represents the total
  number of observables possible in each case.  $\mathcal{P}_{(ij)}$
  indicates the sum $\mathcal{P}_{ij}+\mathcal{P}_{ji}$ and
  $\mathcal{P}^{(\pm)}_i=\mathcal{P}^{+}_i+\mathcal{P}^{-}_i$, with
  identical definitions for the $\mathcal{A}$'s.}
\label{tab3}
\end{center}
\end{table}

Finally, we note that in some of these scenarios, it will be possible
to extract the value of $m_b$. This is advantageous for two reasons.
First, it permits a direct comparison with the theoretical estimates
of $m_b$ \cite{luke}. Second, for some measurements in the $B$ system,
it is necessary to input $m_b$ from theory, which increases the
systematic (theoretical) uncertainty of the measurement. By contrast,
we see that the double-spin analysis of $b\to s \tau^+ \tau^-$ will
not suffer from this type of systematic error.

\section{Conclusions}

In the standard model (SM), the inclusive decay $b\to s \tau^+ \tau^-$
is described by five theoretical parameters: $C_7^{eff}$, $C_{10}$,
${\rm Re}(C_9^{eff})(\hat{s})$, ${\rm Im}(C_9^{eff})(\hat{s})$ and
$m_b$, where $\hat{s}$ is related to the momentum transferred to the
lepton pair. We would like to be able to test this description.

In this paper, we have calculated all single- and double-spin
asymmetries in the decay $b\to s \tau^+ \tau^-$. We have shown that
there are many different ways of testing the SM description of this
decay. In all, there are a total of 31 different asymmetries. However,
only 9 of these are predicted to be measurable, i.e.\ have values
larger than 10\%. (Indeed some asymmetries are expected to vanish in
the SM.) Should any of the small asymmetries be found to have large
values, this would be a clear signal of new physics (NP). Furthermore,
the SM predicts certain relationships among the asymmetries when the
spins ${\bf s^{+}}$ and ${\bf s^{-}}$ are interchanged. Should these
relations be violated, this would also indicate the presence of NP. In
fact, this could give us some clue as to whether the NP is
CP-conserving or CP-violating.

Apart from these signals of NP, whether or not the SM can be tested
depends crucially on which types of measurements can be made. For
example, if one cannot perform $b$-tagging, and can measure only a
single individual $\tau$ spin, then there are only two sizeable
observables. This is not enough to test the SM. On the other hand, if
one can measure both $\tau$ spins, but cannot tag the flavor of the
$b$, then there are a total of five measurable observables. This is
enough to determine the theoretical unknowns, but does not provide the
necessary redundancy to test the SM.

On the other hand, if one can perform $b$-tagging, but can only
measure a single $\tau$ spin, then there are 7 sizeable observables.
This can provide a redundant test of the SM. The optimal scenario is
if $b$-tagging is possible, and one can measure the polarizations of
both the $\tau^+$ and $\tau^-$. In this case, there are a total of 10
independent measurements, which would greatly overconstrain the SM. If
new physics is not found, this would precisely determine the five
theoretical parameters.

Note that testing the SM implies that the quantity $m_b$ will be
extracted from the experimental data. This will allow us to compare
the experimental value of $m_b$ with the theoretical estimates of this
same quantity. Furthermore, as the measurements do not rely on
theoretical input, the systematic error will be correspondingly
reduced.

\bigskip
\noindent
{\bf Acknowledgements}:
N.S. and R.S. thank D.L. for the hospitality of the Universit\'e de
Montr\'eal, where part of this work was done. The work of D.L. was
financially supported by NSERC of Canada. The work of Nita Sinha was
supported by a project of the Department of Science and Technology,
India, under the young scientist scheme.

\section{Appendix}

In this Appendix, we calculate the square of the amplitude in
Eq.~(\ref{amplitude}), keeping the spins of both final-state leptons.
We define $p_b$, $p_s$, $p_+$ and $p_-$ to be the momenta of the
$b$-quark, $s$-quark, $\tau^+$ and $\tau^-$, respectively, with $q =
p_b - p_s = p_+ + p_-$. The spins of the $\tau^+$ and $\tau^-$ are
denoted by $s_+$ and $s_-$, respectively. We have
\beq
{|\cal M|}^2 = |T_9|^2 + |T_{10}|^2 + |T_7|^2 + 2 {\rm Re}
\left(T_9^\dagger T_{10}\right) + 2 {\rm Re}\left(T_9^\dagger
T_7\right) + 2 {\rm Re}\left(T_{10}^\dagger T_7\right) ~.
\eeq
Summing over the $s$-quark spin and averaging over the $b$-quark spin,
We find
\bea
{1\over 2} \sum_{ b,s \ spins} |T_9|^2 & = & {\alpha^2 G_F^2 \over
\pi^2} |V_{tb} V_{ts}^*|^2 |C_9^{eff}|^2 \nn \\
& \times & \left\{ {(m_b^2 - q^2)\over 2} \left( - p_- \cdot s_+ \, p_+
\cdot s_- + {q^2\over 2}s_+ \cdot s_- + m_\tau^2 (1- s_+ \cdot s_-) \right)
\right. \nn\\
&& + (1- s_+ \cdot s_-) \left( p_b \cdot p_+ \, p_s \cdot p_- + p_s
\cdot p_+ \, p_b \cdot p_- \right) \nn\\
&&- {q^2\over 2} \left[ p_b \cdot s_+ \, p_s \cdot s_- + p_s \cdot s_+
\, p_b \cdot s_-\right] \nn\\
&& + s_+ \cdot p_- \left[ p_b \cdot p_+ \, p_s \cdot s_- + p_s \cdot
p_+ \, p_b \cdot s_- \right] \nn\\
&& + s_- \cdot p_+ \left[ p_b \cdot p_- \, p_s \cdot s_+ + p_s \cdot
p_- \, p_b \cdot s_+ \right] \nn\\
&& +m_\tau \bigg[ p_s \cdot(p_+ + p_-) \, p_b \cdot (s_+ +
s_-) \nn\\
&& \hskip15truemm - p_b \cdot (p_+ + p_-) \, p_s \cdot (s_+ + s_-)
\bigg] \Bigg\} ~,
\eea   
\bea
{1\over 2} \sum_{ b,s \ spins} |T_{10}|^2 & = & {\alpha^2 G_F^2 \over
\pi^2} |V_{tb} V_{ts}^*|^2 |C_{10}|^2 \nn \\
& \times & \left\{ -{(m_b^2 - q^2)\over 2} \left( - p_- \cdot s_+ \, p_+
\cdot s_- + {q^2\over 2} s_+ \cdot s_- + m_\tau^2 (1- s_+ \cdot s_-)
\right) \right. ~ \nn\\
&& + (1+ s_+ \cdot s_-) \left( p_b \cdot p_+ \, p_s \cdot p_- + p_s
\cdot p_+ \, p_b \cdot p_- \right) ~ \nn\\
&&- \left( 2m_\tau^2 - {q^2\over 2} \right) \left[ p_b \cdot s_+ \, p_s
\cdot s_- + p_s \cdot s_+ \, p_b \cdot s_-\right] ~ \nn\\
&& - s_+ \cdot p_- \left[ p_b \cdot p_+ \, p_s \cdot s_- + p_s \cdot
p_+ \, p_b \cdot s_- \right] ~ \nn\\
&& - s_- \cdot p_+ \left[ p_b \cdot p_- \, p_s \cdot s_+ + p_s \cdot
 p_- \, p_b \cdot s_+ \right] ~ \nn\\
&& -m_\tau \bigg[ p_s \cdot (p_+ - p_-) \, p_b \cdot (s_+
- s_-) \nn\\
&& \hskip15truemm - p_b \cdot (p_+ - p_-) \, p_s \cdot (s_+ - s_-)
\bigg] \Bigg\} ~,
\eea

\bea
\sum_{ b,s \ spins} {\rm Re}\Big[T_9^\dagger T_{10}\Big] & = & {\alpha^2 G_F^2
\over \pi^2} |V_{tb} V_{ts}^*|^2  \nn \\
& \times & \Bigg\{ 2 {\rm Re}(C_9^{eff} C_{10}^\ast )\bigg[ m_\tau^2 \big[
p_s \cdot s_- \, p_b \cdot s_+ - p_b \cdot s_- \, p_s \cdot s_+ \big]
\nn\\
&& \hskip5truemm - {q^2\over 2}(p_s \cdot p_- - p_s \cdot p_+) ~ \nn
\\
&& \hskip5truemm + m_\tau\big[ p_b \cdot p_+ \, p_s \cdot s_- + p_s \cdot
p_+ \, p_b \cdot s_- \nn\\
&& \hskip5truemm - p_b \cdot p_- \, p_s \cdot s_+ - p_s \cdot p_- \,
p_b \cdot s_+\big] \bigg] ~ \nn \\
&& +{\rm Im}(C_9^{eff}C_{10}^\ast ){\epsilon}_{\mu\alpha\beta\phi}
\bigg[ \Big[2m_\tau + (p_s+p_b) \cdot s_+ \Big] p_s^\mu p_-^\alpha
s_-^\beta p_+^\phi ~ \nn \\
&& \hskip5truemm -\Big[ 2m_\tau + (p_s+p_b) \cdot s_- \Big] p_s^\mu
 p_+^\alpha s_+^\beta p_-^\phi~ \nn \\
&& \hskip5truemm - (p_s+p_b) \cdot p_+ \, p_s^\mu p_-^\alpha s_-^\beta
s_+^\phi + (p_s+p_b) \cdot p_- \, p_s^\mu p_+^\alpha s_+^\beta
s_-^\phi ~ \nn \\
&& \hskip5truemm + (p_- - p_+) \cdot p_s \, p_-^\mu p_+^\alpha
s_+^\beta s_-^\phi \bigg] \Bigg\} ~,
\label{T9T10}
\eea

\newpage

\bea
{1\over 2} \sum_{ b,s \ spins} |T_7|^2 & = & {\alpha^2 G_F^2 \over
\pi^2} |V_{tb} V_{ts}^*|^2 { m_b^2\over q^4} |C_{7}|^2 \nn \\
& \times & \Bigg\{4 m_b^2 m_\tau\bigg[ p_s \cdot (p_+ + p_-) \, q
\cdot (s_- + s_+) - q^2 \, p_s \cdot (s_- + s_+) \bigg] ~ \nn \\
&& +2m_\tau^2 m_b^2 (m_b^2-q^2)\big[(1- s_+ \cdot s_- \big] +q^2 m_b^2
(m_b^2-q^2) ~ \nn \\
&&-2q^2 \bigg[2 \big[ p_s \cdot p_- \, p_b \cdot p_+ + p_s \cdot p_+
\, p_b \cdot p_- \big]-q^2 \, p_b \cdot p_s \bigg]\big[(1- s_+ \cdot
s_-\big] ~ \nn \\
&&-4q^2 \bigg[ s_+ \cdot p_- \big[ p_s \cdot s_- \, p_b \cdot p_+ +
p_b \cdot s_- \, p_s \cdot p_+ \big] \nn\\
&& \hskip10truemm + s_- \cdot p_+ \big[ p_s \cdot p_- \, p_b \cdot s_+
+ p_b \cdot p_- \, p_s \cdot s_+ \big] \bigg] ~ \nn \\
&& + 2q^2(m_b^2-q^2) \, s_+ \cdot p_- \, s_- \cdot p_+ \nn\\
&& + 2q^4\big[ p_s \cdot s_- \, p_b \cdot s_+ + p_b \cdot s_- \, p_s
\cdot s_+ \big] \Bigg\} ~,
\label{T7squared}
\eea
\bea
\sum_{ b,s \ spins} {\rm Re}\Big[ T_9^\dagger T_7\Big] & = & {\alpha^2 G_F^2
\over \pi^2} |V_{tb} V_{ts}^*|^2 {m_b^2\over q^2} \nn \\
& \times & \Bigg\{ {\rm Re}(C_9^{eff}c_{7}^\ast )\bigg[ (m_b^2-q^2)
\big[ q^2+2m_\tau^2 -2m_\tau^2 \, s_+ \cdot s_- \big] \nn\\
&& -4 m_\tau \big[ p_b \cdot (p_+ + p_-) \, p_s \cdot (s_+ + s_-) \nn\\
&& \hskip10truemm - p_s \cdot (p_+ + p_-) \, p_b \cdot (s_+ + s_-)
\big] \bigg] \Bigg\} ~,
\eea
\bea
\sum_{ b,s \ spins} {\rm Re}\Big[ T_{10}^\dagger T_7\Big] & = &
{\alpha^2 G_F^2 \over \pi^2} |V_{tb} V_{ts}^*|^2 {m_b^2\over q^2} ~
\nn \\
& \times & \Bigg\{ -2 {\rm Re}(C_{10}C_7^{eff^*} ) \bigg[
  -{m_\tau(m_b^2-q^2)\over 2} \big[p_+ \cdot s_- - p_- \cdot s_+\big]
  ~ \nn \\
&& + q^2 \big[ p_s \cdot p_- - p_s \cdot p_+ \big] - 2m_\tau^2 \big[
p_s \cdot s_- \, p_- \cdot s_+ - p_+ \cdot s_- \, p_s \cdot s_+ \big]
~ \nn \\
&& + m_\tau \, q^2 p_s \cdot (s_+ - s_-) + m_\tau \, p_s \cdot (p_- -
p_+)(s_- \cdot p_+ + s_+ \cdot p_- ) \bigg] ~ \nn \\
&& -4 {\rm Im}(C_{10}C_7^{eff^*} ){\epsilon}_{\alpha\beta\mu\rho}
\bigg[ - m_\tau \, p_s^\alpha p_+^\beta s_+^\mu p_b^\rho + m_\tau
\, p_s^\alpha p_-^\beta s_-^\mu p_b^\rho \nn\\
&& \qquad + m_\tau^2 \, s_-^\alpha p_s^\beta s_+^\mu p_b^\rho \bigg]
\Bigg\} ~ .
\label{T7T10}
\eea
In the above, we have used the convention $\epsilon_{0123} = +1$.
Note that, as mentioned in Sec.~2, $C_7^{eff}$ and $C_{10}$ are
expected to be real; only $C_9^{eff}$ is complex. However, in the
expressions above, for completeness we have included both real and
imaginary pieces of all Wilson coefficients.



\begin{thebibliography}{99}

\bibitem{bsllSM} W.~S.~Hou, R.~S.~Willey and A.~Soni, Phys.\ Rev.\
  Lett.\ {\bf 58}, 1608 (1987) [Erratum-ibid.\ {\bf 60}, 2337 (1988)];
  N.~G.~Deshpande and J.~Trampetic, Phys.\ Rev.\ Lett.\ {\bf 60}, 2583
  (1988); A.~Ali, T.~Mannel and T.~Morozumi, Phys.\ Lett.\ B {\bf
  273}, 505 (1991); A.~F.~Falk, M.~E.~Luke and M.~J.~Savage, Phys.\
  Rev.\ D {\bf 49}, 3367 (1994); A.~Ali, G.~F.~Giudice and T.~Mannel,
  \zpc{67}{1995}{417}; C.~Greub, A.~Ioannisian and D.~Wyler, Phys.\
  Lett.\ B {\bf 346}, 149 (1995); J.~L.~Hewett, \prd{53}{1996}{4964};
  F.~Kruger and L.~M.~Sehgal, \plb{380}{1996}{199}; A.~Ali, G.~Hiller,
  L.~T.~Handoko and T.~Morozumi, Phys.\ Rev.\ D {\bf 55}, 4105 (1997);
  C.~S.~Kim, T.~Morozumi and A.~I.~Sanda, Phys.\ Rev.\ D {\bf 56},
  7240 (1997); T.~M.~Aliev, C.~S.~Kim and M.~Savci, Phys.\ Lett.\ B
  {\bf 441}, 410 (1998); C.~Q.~Geng and C.~P.~Kao, Phys.\ Rev.\ D {\bf
  57}, 4479 (1998); S.~Fukae, C.~S.~Kim, T.~Morozumi and T.~Yoshikawa,
  Phys.\ Rev.\ D {\bf 59}, 074013 (1999); Y.~G.~Kim, P.~Ko and
  J.~S.~Lee, Nucl.\ Phys.\ B {\bf 544}, 64 (1999); S.~Fukae, C.~S.~Kim
  and T.~Yoshikawa, \newprd{61}{2000}{074015}; M.~Zhong, Y.~L.~Wu and
  W.~Y.~Wang, arXiv:hep-ph/0206013; A.~Ghinculov, T.~Hurth, G.~Isidori
  and Y.~P.~Yao, arXiv:hep-ph/0208088; H.~M.~Asatrian, K.~Bieri,
  C.~Greub and A.~Hovhannisyan, arXiv:hep-ph/0209006.

\bibitem{bsllNP} P.~L.~Cho, M.~Misiak and D.~Wyler, Phys.\ Rev.\ D
  {\bf 54}, 3329 (1996); Y.~Grossman, Z.~Ligeti and E.~Nardi, Phys.\
  Rev.\ D {\bf 55}, 2768 (1997); J.~L.~Hewett and J.~D.~Wells, Phys.\
  Rev.\ D {\bf 55}, 5549 (1997); T.~Goto, Y.~Okada, Y.~Shimizu and
  M.~Tanaka, Phys.\ Rev.\ D {\bf 55}, 4273 (1997) [Erratum-ibid.\ D
  {\bf 66}, 019901 (2002)]; L.~T.~Handoko, Nuovo Cim.\ A {\bf 111}, 95
  (1998); J.~H.~Jang, Y.~G.~Kim and J.~S.~Lee, Phys.\ Rev.\ D {\bf
  58}, 035006 (1998); T.~G.~Rizzo, Phys.\ Rev.\ D {\bf 58}, 114014
  (1998); S.~Rai Choudhury, A.~Gupta and N.~Gaur, Phys.\ Rev.\ D {\bf
  60}, 115004 (1999); C.-S.~Huang and S.~H.~Zhu, Phys.\ Rev.\ D {\bf
  61}, 015011 (2000) [Erratum-ibid.\ D {\bf 61}, 119903 (2000)].

\bibitem{AGM} A.~Ali, G.~F.~Giudice and T.~Mannel, Ref.~\cite{bsllSM}.

\bibitem{Hewett} J.~L.~Hewett, Ref.~\cite{bsllSM}.

\bibitem{KS} F.~Kruger and L.~M.~Sehgal, Ref.~\cite{bsllSM}.

\bibitem{smallPNref} Refs.~\cite{KS} and \cite{FKY} give $\langle P_N
  \rangle_\tau = 0.05$ and $\langle P_N \rangle_\tau = 0.02$,
  respectively. The authors of Ref.~\cite{FKY} cut out resonances
  below the $\Psi'$, which increases the asymmetry, while the authors
  of Ref.~\cite{KS} have used a cut of $\pm 30$ MeV for the $\Psi'$
  resonances. As we show later in the paper, we find $\langle P_N
  \rangle_\tau = 0.014$.

\bibitem{FKY} S.~Fukae, C.~S.~Kim and T.~Yoshikawa, Ref.~\cite{bsllSM}.

\bibitem{luke} A.X. El-Khadra and M. Luke, hep-ph/0208114.

\bibitem{hamiltonian} B. Grinstein, M.~J. Savage and M.~B. Wise,
  \npb{319}{1989}{271}; A.~J. Buras and M. Munz, \prd{52}{1995}{186};
  M.~Misiak, Nucl.\ Phys.\ B {\bf 393}, 23 (1993) [Erratum-ibid.\ B
  {\bf 439}, 461 (1995)].

\bibitem{resonances} C.~S. Lim, T. Morozumi and A.~I. Sanda,
  \plb{218}{1989}{343}; N.~G. Deshpande, J. Trampeti\'c and K.~Panose,
  \prd{39}{1989}{1461}; P.~J. O'Donnell, M.~Sutherland and
  H.~K.~K.~Tung, \prd{46}{1992}{4091}; P.~J.~O'Donnell and
  H.~K.~K.~Tung, \prd{43}{1991}{R2067}; F.~Kruger and L.~M. Sehgal,
  Ref.~\cite{bsllSM}.

\bibitem{Ali} A.~Ali, T.~Mannel and T.~Morozumi, Ref.~\cite{bsllSM}.

\bibitem{CPTref} {\it Particle Physics and Introduction to Field
  Theory}, by T.~D.~Lee, Harwood Academic Publishers (1981).

\end{thebibliography}
\end{document}